\documentclass[amsmath,amssymb,aps,prb,showkeys, floatfix,10pt,superscriptaddress, twocolumn]{revtex4-2}

\usepackage{mathtools}
\usepackage[caption = false]{subfig}
\usepackage{graphicx}
\usepackage{standalone}
\usepackage{dcolumn}
\usepackage{bm}
\usepackage{xcolor}
\usepackage{physics}
\usepackage{enumitem}
\usepackage{calligra}
\usepackage[colorlinks=true, linkcolor=red, citecolor=blue, urlcolor=blue]{hyperref}
\usepackage[sort&compress]{natbib}
\usepackage{orcidlink}
\usepackage[normalem]{ulem}
\usepackage{lineno}
\usepackage{kotex}
\usepackage{xfrac}

\DeclareGraphicsExtensions{.pdf,.eps,.png,.jpg,.mps}

\makeatletter
\renewcommand*{\fnum@figure}{{\normalfont\bfseries \figurename~\thefigure}}
\renewcommand*{\@caption@fignum@sep}{\textbf{:}}
\makeatother

\newcommand{\apctpadd}{Asia Pacific Center for Theoretical Physics, Pohang, Gyeongbuk, 37673, Republic of Korea}
\newcommand{\postechadd}{Department of Physics, Pohang University of Science and Technology, Pohang, Gyeongbuk 37673, Republic of Korea}

\begin{document}
% \linenumbers

% \title{Spectral and dynamical properties of discrete time quasicrystals driven by quasiperiodic Fibonacci kicks of a staggered periodic potential}
%\title{Spectral and dynamical properties of quasiperiodically-kicked staggered periodic potential chains}
% \title{Anomalous quasiperiodic Rabi oscillations driven by quasiperiodic kicking of a periodic staggered potential }
% Spectral and dynamical properties of a tight-binding chain driven by quasiperiodic kicks of a staggered periodic potential
\title{Spectral and dynamical properties of quasiperiodically kicked staggered chains}

\author{Sanghoon Lee\orcidlink{0009-0001-2248-5534}}
%\email{sanghoon.lee@apctp.org}
\affiliation{\apctpadd}

\author{Kyoung-Min Kim\orcidlink{0000-0003-2468-3152}}
\email{kyoungmin.kim@apctp.org}
\affiliation{\apctpadd}
\affiliation{\postechadd}

\date{\today}

\begin{abstract}
% Problem/Question :-
We study a one-dimensional lattice driven by Fibonacci-ordered kicks of a staggered potential.
We ask what the quasienergy spectra of periodically repeated finite segments reveal about the nonrepeating drive.
% Main result :-
% > Claim 1 : what approximant spectra capture
For weak kicks, one temporal Fourier amplitude predicts both the quasienergy gap and the frequency of interband oscillations under periodic repetition.
As the segment grows, the gap tends to zero, but a finite limiting coupling sets the oscillation frequency under the nonrepeating drive.
% > Claim 2 : what approximant spectra miss
At special momenta, the difference between kick counts on even and odd steps determines the exact interband population at any kick strength.
Under the nonrepeating drive, the maximum magnitude of this imbalance grows logarithmically with time.
For weak kicks, the time to first reach a fixed upper-band population therefore grows exponentially in the inverse kick strength.
Under repetition of a fixed even-length segment, it grows linearly in the inverse kick strength.
% Broader perspective
Temporal ordering thus controls interband oscillations and the time needed to populate the upper band.
\end{abstract}


\maketitle

\section{Introduction}
\label{sec:introduction}
 
% 1. Basic Introduction
% 2. More detailed background
% 3. General problem
% 4. Here we show
% 5. Table of contents
 
% 1. Basic Introduction
Periodic driving controls quantum dynamics and quasienergy spectra~\cite{oka2019floquet,eckardt2017colloquium}.
When the drive is a sequence of pulses, each pulse mixes probability amplitudes, and free evolution between pulses changes their relative phases.
For equally spaced identical pulses in a two-level system, the population dynamics shows interference analogous to grating diffraction~\cite{vitanov1995coherent}.
If the pulses instead follow a deterministic nonrepeating sequence such as Fibonacci order, there is no single drive period.
Such a drive is often described by two incommensurate frequencies in a quasi-Floquet framework~\cite{ho1983semiclassical,verdeny2016quasi}.
%In this framework, Fourier components of the drive set the resonance strengths.
 
% 2. More detailed background
% > 2-1
Quasiperiodicity can enter through the spatial structure, the driving sequence, or both.
Spatial quasiperiodicity alone affects spectra and localization in undriven lattice models~\cite{aubry1980analyticity,biddle2010predicted,ganeshan2015nearest,goncalves2023critical,lee2023criticala,lee2023criticalb,lee2026structural,lee2026mobility}, including Fibonacci chains~\cite{kohmoto1983localization,ostlund1983one,kohmoto1987critical,jagannathan2021the,tanese2014fractal}.
Quasiperiodically kicked Aubry--Andr\'e--Harper lattices, which combine both, can show multifractality and long-lived prethermal behavior~\cite{chen2024multifractality}.
In these lattices, kick timing also matters: random variations in the time between kicks can destroy dynamical localization~\cite{cadez2017dynamical}.

% > 2-2
The role of temporal ordering can be isolated in spatially periodic lattices.
In interacting spin chains, structured aperiodic drives can stabilize long-lived prethermal phases and suppress heating, with time scales set by the drive structure~\cite{zhao2021random,mori2021rigorous}.
Without interactions, Fibonacci-ordered driving of free-fermion chains leads to nonequilibrium steady states~\cite{maity2019fibonacci,nandy2018steady}.
% The role of temporal ordering can be isolated in spatially periodic lattices.
% In interacting lattices, structured aperiodic drives can stabilize long-lived prethermal phases and suppress heating, with time scales set by the drive structure~\cite{else2020long,zhao2021random,mori2021rigorous}.
% Without interactions, Fibonacci-ordered driving can produce dynamical localization in a clean tight-binding chain~\cite{tiwari2024dynamical} and nearly stationary local observables in an integrable spin chain~\cite{maity2019fibonacci}.

% 2-3. Fibonacci-driven two-level dynamics
A drive that preserves the lattice period conserves momentum.
In a noninteracting two-band lattice, each momentum sector forms an independent two-level system.
A Fibonacci word specifies a finite sequence of two types of evolution steps.
Each word concatenates the previous two.
Their propagators obey the same recursion as transfer matrices in a spatial Fibonacci chain~\cite{kohmoto1983localization,sutherland1986simple,ray2019dynamics, luck1988on}.

% 3. General problem :-
This recursion makes finite words the natural route to the nonrepeating drive, which follows the infinite Fibonacci sequence.
A commensurate approximant repeats the same Fibonacci word, and its quasienergy spectrum can be computed using Floquet theory.
The first question is how its quasienergy gaps relate to interband population.
As the word grows, the Floquet zone shrinks to zero, so it is unclear whether this relation survives.
The nonrepeating drive also raises a question about time scales.
At a solvable point of a Fibonacci-driven conformal field theory, entanglement entropy oscillates at word ends, while its envelope grows logarithmically in time~\cite{wen2021periodically}.
This slow growth motivates a second question: how the time to first reach a fixed upper-band population differs between the nonrepeating drive and periodic repetition.

% 4. Here we show: main claims
Here, we answer both questions for a tight-binding chain driven by Fibonacci-ordered kicks of a staggered potential.
The staggered potential doubles the unit cell, giving an independent two-level system at each momentum.
%
% 4-1. Claim 1
First, for a fixed word and weak kicks, a single temporal Fourier amplitude of the kick sequence predicts both the quasienergy gap and the interband Rabi period under repetition.
When this amplitude grows in proportion to word length, the predicted splitting before quasienergy folding approaches a finite value.
The exact gap, in contrast, vanishes as the Floquet zone shrinks.
In the weak-kick limit, this limiting first-order splitting sets the Rabi frequency under the nonrepeating drive.
%
% 4-2. Claim 2
Second, at special momenta, free evolution gives opposite signs to kick contributions on even and odd steps.
The exact interband population at any kick strength is then determined by the kick-count imbalance, the difference between the numbers of kicks on even and odd steps.
Under repetition, the imbalance at word ends grows linearly for even-length words and cancels over pairs of odd-length words.
Under the nonrepeating drive, it remains bounded at word ends, while its maximum magnitude grows logarithmically with time.
For weak kicks, the time to first reach a fixed upper-band population is inversely proportional to the kick strength under repetition of a fixed even-length word.
Under the nonrepeating drive, this time grows exponentially in the inverse kick strength.
%
% 4-3. Overall message
Approximant spectra thus capture the resonant dynamics of the nonrepeating drive but miss the logarithmic accumulation within the words.
% The temporal Fourier amplitude describes resonant interband oscillations, while the kick-count imbalance determines the time needed to populate the upper band.
 
% 5. Table of contents
The paper is organized as follows.
We define the model and observables in Sec.~\ref{sec:model}.
In Sec.~\ref{sec:quasienergy_resonances}, we relate quasienergy gaps to interband oscillations.
In Sec.~\ref{sec:half_integer}, we derive the exact interband population at special momenta and compare repeated words with the nonrepeating drive, followed by conclusions in Sec.~\ref{sec:discussions}.

% SECTION : MODEL
\section{Model}\label{sec:model}
 
\begin{figure}[t]
    \centering
    \includegraphics[width=\linewidth]{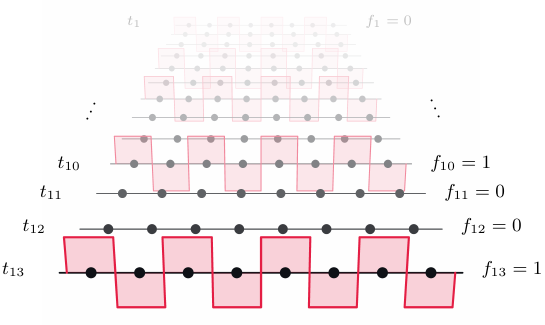}
	\caption{
	\textbf{Fibonacci-ordered kicking.}
	% 1. What is shown :- (a)
	Schematic of a one-dimensional tight-binding chain driven by Fibonacci-ordered kicks.
	% 2. How to read it:
	Time advances from back to front, with earlier snapshots faded for clarity.
	Dots and connecting lines denote lattice sites and nearest-neighbor hopping, respectively.
	% 3. Conditions and definitions:
	Red profiles indicate the alternating on-site kick amplitudes \(\lambda(-1)^\ell\) at steps with \(f_{n}=1\); no kick is applied when \(f_{n}=0\).
	}
    \label{fig:fig1}
\end{figure}

% Define the model
We consider a one-dimensional tight-binding chain driven by instantaneous kicks from a staggered on-site potential.
The Hamiltonian is
\begin{gather}\label{eq:model}
\begin{gathered}
	H(t)  = H_{0} + \sum_{n\in\mathbb{Z}} f_{n} \delta(t - t_{n}) V \\[3pt]
    H_{0} = -J\sum_{\ell\in\mathbb{Z}}\hat{c}_{\ell+1}^{\dagger}\hat{c}_{\ell} + \!\mathrm{h.c.} , \ \ 
    V     = \lambda\!\sum_{\ell\in\mathbb{Z}}(-1)^{\ell} \hat{c}_{\ell}^{\dagger} \hat{c}_{\ell}.
\end{gathered}
\end{gather}
Here, \(J>0\) is the hopping amplitude, and \(T\) is the duration of one time step.
At \(t_n=nT\), a kick is applied if \(f_n=1\) and omitted if \(f_n=0\).
A schematic of the model is shown in Fig.~\ref{fig:fig1}.
The kick sequence \(f_{n}\) follows Fibonacci words \(w_{m}\) of length \(L_{m}\).
Starting from the words \(w_{0} = 1\) and \(w_{1} = 0\), we form each new word by concatenating the two preceding words, with the more recent word first: \(w_{m+1}=w_{m}w_{m-1}\)~\cite{kohmoto1983localization,ostlund1983one}.
For example, \(w_{3} = 010\) describes three time steps, with only the middle kick active.
Under this convention, the word \(w_m\) contains \(L_{m-2}\) active kicks for \(m\geq2\).
The active-kick fraction \(L_{m-2}/L_m\) therefore approaches \(\phi^{-2}\) as \(m\to\infty\), where \(\phi=(1+\sqrt{5})/2\) is the golden ratio.
Since each word \(w_m\) with \(m\geq 1\) is a prefix of \(w_{m+1}\), the words converge to an infinite Fibonacci word, which we call the nonrepeating drive.
We also set \(\hbar=1\) throughout, and \(J = T = 1\).
 
% Define eigenvalue problem
The staggered potential alternates between neighboring sites, so we group pairs of sites into a two-site unit cell and use the reduced momentum \(k\in(-\pi,\pi]\).
At each \(k\), the free Bloch Hamiltonian has bonding and antibonding eigenstates, \(\ket{+,k}\) and \(\ket{-,k}\), respectively, satisfying
\begin{gather}
    H_{0}(k)\ket{\pm,k}=\pm\epsilon(k)\ket{\pm,k}, \quad \epsilon(k)=-2J\cos\frac{k}{2}.
\end{gather}
With a suitable choice of the relative phase between these states, the two operators take the form \(H_{0}(k)=\epsilon(k)\tau_z\) and \(V(k)=\lambda\tau_x\), where \(\tau_{x,y,z}\) are Pauli matrices in the two-band basis.
Thus, each momentum sector evolves as an independent two-level system.
Each step consists of free evolution, which rotates the relative phase of the two bands, followed by a kick.
The free evolution operator and the kick operator are denoted as follows,
\begin{gather}\label{eq:step_operators}
    F(k) = \exp\left[-i\epsilon(k)T\tau_z\right], \quad K_n = \exp\left[-if_n\lambda\tau_x\right],
\end{gather}
and the step operator is \(U_{n}(k) = K_nF(k)\).
An active kick mixes the bands, whereas \(f_n=0\) gives \(K_n=\mathbb I\).
Periodic repetition of the Fibonacci word \(w_m\) defines the \(m\)-th commensurate approximant.
The resulting drive has a period of \(\mathcal{T}_m = L_{m}T\).
We define the Floquet eigenstates \(\ket{\phi_{\alpha,k}}\) and their quasienergies \(E_{\alpha}(k)\) by
\begin{gather}\label{eq:diagonalization}
\begin{gathered}
	\mathcal{U}(\mathcal{T}_{m}, k)\ket{\phi_{\alpha, k}} = \exp(-iE_{\alpha}(k) \mathcal{T}_{m})\ket{\phi_{\alpha, k}},\\[3pt]
	\mathcal{U}(t_{n}, k) = U_{n}(k)\cdots U_{2}(k)U_{1}(k),
\end{gathered}
\end{gather}
where \(\mathcal{U}(t_n,k)\) describes evolution through the first \(n\) steps, and \(\alpha\) labels the two Floquet bands.
The quasienergies are defined modulo \(\Omega_m\) and chosen within the first quasienergy Brillouin zone,
\begin{gather}
    E_\alpha(k)\in(-\Omega_m/2,\Omega_m/2],
    \qquad \Omega_m=\frac{2\pi}{\mathcal{T}_m}.
\end{gather}
 
% Define the main observable:
To relate the Floquet spectrum to population dynamics, we start from the lower free-band state \(\ket{+,k}\).
The upper-band population after \(n\) steps is
\begin{gather}\label{eq:population}
    P(t_n,k) = \left| \bra{-,k}\mathcal U(t_n,k)\ket{+,k} \right|^{2}.
\end{gather}
To characterize the temporal harmonics of the \(m\)-th approximant, we use the finite-word Fourier amplitude
\begin{gather}\label{eq:fibonacci_structure_factor}
    \mathcal{S}(\omega,\mathcal{T}_m) = \sum_{j=1}^{L_m}f_j\exp(-i\omega t_j).
\end{gather}
Its squared magnitude gives the spectral intensity of the finite kick sequence.
% Its squared magnitude gives the spectral intensity of the kick sequence, analogous to the role of the filter function of a pulse sequence in dynamical decoupling~\cite{cywinski2008how}.
% Symmetries of the model:
Due to symmetries, we consider \(0\leq\lambda\leq\pi/2\) (Appendix~\ref{app:symmetries}).

% SECTION : QUASIENERGY GAPS AND INTERBAND OSCILLATIONS
\section{Quasienergy gaps and interband oscillations}
\label{sec:quasienergy_resonances}
 
% > Section overview
% 1 What the section is about :-
In this section, we compare the quasienergy gaps of the approximants with the interband dynamics for weak kicks.
We first identify the crossings of the free quasienergy bands at which weak kicks open gaps.
We then compare the gaps and the population oscillations at these crossings with a first-order prediction.
Finally, we examine how this comparison changes as the word grows and under the nonrepeating drive.
 
\begin{figure}[t]
    \centering
    \includegraphics[width=\linewidth]{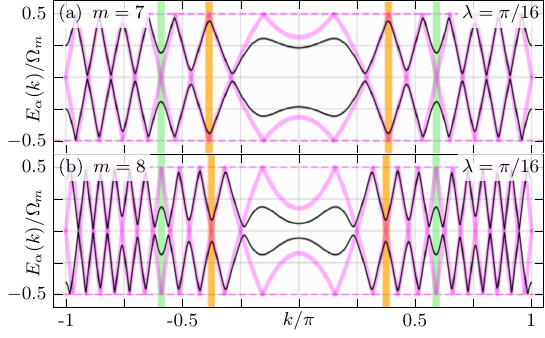}
	\caption{
	\textbf{Quasienergy bands.}
	(a),(b) Free and driven quasienergy spectra for Fibonacci approximants \(m=7,8\), respectively.
	% 2. How to read it:
	Magenta and black curves show the zone-folded free quasienergy spectra and driven spectra, respectively.
	Green and orange strips mark free-band crossing momenta for Fibonacci harmonics \(b=+1\) and \(b=+4\) [Eq.~\eqref{eq:temporal_bragg_condition}], respectively; these momenta depend on \(m\).
	Horizontal magenta dashed lines mark the Floquet-zone boundaries.
	% 3. Conditions and definitions:
	The driven spectra use \(\lambda=\pi/16\), \(\Omega_m=2\pi/\mathcal{T}_m\) and \(J=T=1\).
	}
    \label{fig:fig2}
\end{figure}
 
% > Free quasienergy bands (magenta)
% 1 What the figure is about :-
The magenta curves in Figs.~\ref{fig:fig2}(a) and (b) show the free quasienergy spectra (\(\lambda=0\)) for the \(m=7,8\) Fibonacci approximants, respectively.
% >> Folding
% 2 What the key pattern is (build up) :-
The two branches cross repeatedly at the zone center, \(E/\Omega_{m}=0\), and at the equivalent zone boundaries, \(E/\Omega_{m}=\pm1/2\).
The crossings occur in pairs at opposite momenta and become more closely spaced toward \(k=\pm\pi\).
% 3 What meaning the pattern has (compared to what) :-
These crossings arise from folding the free bands into the Floquet zone.
Without kicks, the evolution over one word is \(\mathcal{U}(\mathcal{T}_m,k)=\exp[-i\epsilon(k)\mathcal{T}_m\tau_z]\).
The two free-band states acquire the same phase after one word at momenta \(k=k_p\) satisfying
\begin{gather}\label{eq:folded_crossing_condition}
    |\epsilon(k_p)|\mathcal{T}_m=p\pi,
    \qquad
    p=1,\ldots,\left\lfloor\frac{2J\mathcal{T}_m}{\pi}\right\rfloor.
\end{gather}
As the word length \(L_m\) increases, the Floquet zone narrows and more folded crossings appear.
At each crossing,
\(\mathcal{U}(\mathcal{T}_m,k_p)=(-1)^p\mathbb{I}\),
so even \(p\) gives a crossing at the zone center, while odd \(p\) gives one at the zone boundary.
The case \(p=0\) corresponds to \(k=\pm\pi\); for detailed discussion, see Appendix~\ref{app:exact}.

% 2 Harmonic labels and their relation :-
We next relate the free-band crossings to the temporal Fourier components of the Fibonacci word.
Within a given approximant, \(p\) labels a crossing with free-band separation \(\omega(k_p)=2|\epsilon(k_p)|=p\Omega_m\).
We label the Fibonacci harmonics by an integer \(b\), with frequencies \(b\nu_m\), where \(\nu_m T/2\pi = L_{m-2}/L_m\) is the fraction of active kicks.
Since \(\nu_m\to2\pi\phi^{-2}/T\) as \(m\to\infty\), a fixed \(b\) tracks the same limiting harmonic across approximants.
To match this harmonic to a crossing, we compare \(b\nu_m\) with \(p\Omega_m\).
Frequencies differing by \(2\pi/T\) have the same phase at every kick time \(t_n=nT\).
The resonance condition is therefore
\begin{gather}\label{eq:temporal_bragg_condition}
    p\equiv bL_{m-2}\pmod{L_m}.
\end{gather}
For the harmonics considered below, we choose the solution with \(1\leq p\leq L_m-1\).
If this solution also lies within the range specified in Eq.~\eqref{eq:folded_crossing_condition}, we denote the corresponding momentum by \(k_b\equiv k_p\).
% Therefore, the index \(p\) labels a crossing within a given approximant, whereas \(b\) labels a harmonic across approximants.

% 3 What the figure shows :-
The green and orange shaded strips in Figs.~\ref{fig:fig2}(a) and (b) mark the crossing pairs selected by \(b=+1\) and \(b=+4\), respectively.
The selected momenta change only slightly from \(m=7\) to \(m=8\), consistent with the convergence of \(b\nu_m\) modulo \(2\pi/T\).
% 4 What meaning the pattern has :-
Their quasienergy locations, however, can change with the parity of \(p\).
For \(m=7,8\), Eq.~\eqref{eq:temporal_bragg_condition} gives \(p=8,13\) for \(b=+1\) and \(p=11,18\) for \(b=+4\).
Since even \(p\) lies at the zone center and odd \(p\) at the boundary, the green crossings switch from the center to the boundary, while the orange crossings switch in the opposite direction.

% > Driven spectrum (black)
% 1 What the figure is about :-
We next discuss how the kicks open quasienergy gaps at the free-band crossings and what determines their sizes.
The black curves in Figs.~\ref{fig:fig2}(a) and (b) show the quasienergy spectra under Fibonacci-ordered kicks at \(\lambda=\pi/16\).
% 2 What the key pattern is (build up) :-
The kicks open gaps at the free-band crossings, but they do not affect all of these crossings equally.
At some crossings, the black curves remain close to the magenta curves, whereas at others they separate substantially.
% 3 Interference mechanism :-
The staggered kicks couple the two free-band states.
To first order in \(\lambda\), an interband transition at a selected crossing \(k=k_b\) can occur at any one of the active kicks.
The interband transition amplitude \(a_j\) induced by the active kick at time \(t_j\) is
\begin{align}
    a_{j} &:= \bra{-,k_b} F^{L_m-j}(k_b)\, K_{j}\, F^j(k_b) \ket{+,k_b} \notag\\[3pt]
    &\phantom{:}\simeq -i(-1)^p\lambda \exp\left(i\omega(k_b)t_{j}\right),
\end{align}
where \((-1)^p\) is the common free-evolution phase at the crossing.
For two active kicks at the time steps \(t_{j}\) and \(t_{\ell}\), adding their amplitudes \(a_{j}\) and \(a_{\ell}\) gives
\begin{gather}
    |a_j+a_{\ell}|^2 \simeq |a_j|^2\!+ |a_{\ell}|^2 \!+ 2\lambda^2 \! \cos[\omega(k_b)(t_j-t_{\ell})].
\end{gather}
A positive cosine term enhances the squared amplitude through constructive interference, while a negative term reduces it through destructive interference.
%
% 4 First-order transition amplitude :-
Adding the amplitudes from all active kicks in one word gives
\begin{gather}\label{eq:word_transition_amplitude}
    \bra{-,k_b}\mathcal U(\mathcal{T}_m,k_b)\ket{+,k_b} \!\simeq\! -i(-1)^{p}\!\lambda \mathcal{S}^{*}\!(\omega(k_b),\mathcal{T}_m),
\end{gather}
where \(\mathcal{S}\) is the Fourier amplitude in Eq.~\eqref{eq:fibonacci_structure_factor}.
%
% 5 Gap formula :-
At each crossing, the interband coupling shifts the two degenerate quasienergies in opposite directions.
To first order, their separation before folding into the Floquet zone, which we call the unfolded splitting, is
\begin{gather}\label{eq:unfolded_quasienergy_splitting}
    \varDelta{E}_b(m)\simeq
    \frac{2\lambda}{\mathcal{T}_m}
    \left|\mathcal{S}(\omega(k_b),\mathcal{T}_m)\right|.
\end{gather}
This dependence on the Fourier amplitude explains the unequal gap sizes in Fig.~\ref{fig:fig2}.
 
\begin{figure}[t]
    \centering
    \includegraphics[width=\linewidth]{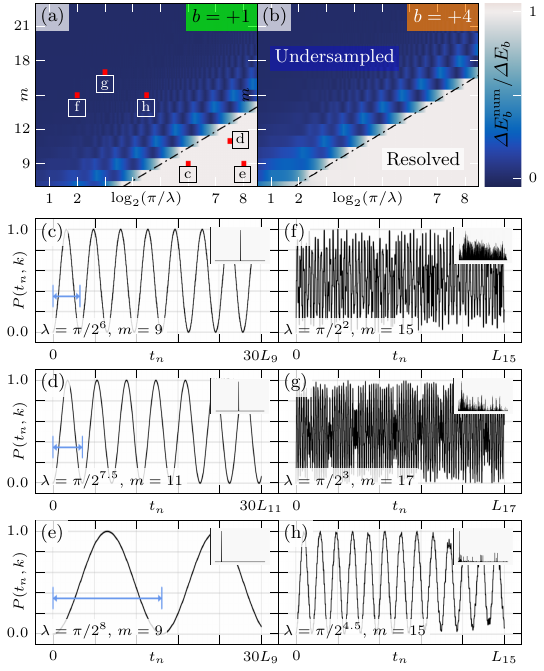}
	\caption{
	\textbf{Resolved and undersampled regimes.}
	% 1. What is shown :- (a),(b)
	(a),(b) Ratio of the exact gap \(\varDelta{E}_{b}^{\rm num}\) to the unfolded splitting \(\varDelta{E}_b\) [Eq.~\eqref{eq:unfolded_quasienergy_splitting}], for \(b=+1\) and \(b=+4\), respectively.
	% 3. Conditions and definitions:
	The exact gap is the shorter separation on the quasienergy circle, obtained by diagonalizing the one-word propagator.
	% 2. How to read it:
	Dash-dotted curves mark \(\lambda|\mathcal{S}|=\pi/2\) [Eq.~\eqref{eq:gap_rabi_boundary}], where the unfolded splitting reaches the maximum exact gap \(\pi/\mathcal{T}_m\).
	Red squares in (a) mark the parameters used in (c)--(h).
	%
	% 1. What is shown :- (c)--(h)
	(c)--(e) Exact kick-resolved upper-band populations over repeated Fibonacci words.
	(f)--(h) Kick-resolved populations within a single word for parameters in the undersampled regime.
	% 2. How to read it:
	Blue arrows in (c)--(e) indicate the weak-kick Rabi periods \(N_b\) [Eq.~\eqref{eq:fibonacci_rabi_probability}].
	Insets show the power spectra of the population traces on a logarithmic scale.
	}
    \label{fig:fig3}
\end{figure}
 
% > Gap maps
% 1 What the figure is about :-
The maps in Figs.~\ref{fig:fig3}(a) and (b) compare the first-order prediction against the exact gaps for \(b=+1\) and \(b=+4\), respectively.
We obtain the exact gap \(\varDelta{E}_{b}^{\rm num}\) by numerically diagonalizing the one-word propagator in Eq.~\eqref{eq:diagonalization}.
The color shows the ratio \(\varDelta{E}_{b}^{\rm num}/\varDelta{E}_{b}\), where \(\varDelta{E}_{b}\) is given by Eq.~\eqref{eq:unfolded_quasienergy_splitting}.
% 2 What the key pattern is (build up) :-
A ratio near unity means that Eq.~\eqref{eq:unfolded_quasienergy_splitting} gives the exact gap.
Weaker kicks lie to the right, and larger \(m\) corresponds to longer words.
For both harmonics, the lower-right region is nearly white, where the exact gap agrees with the first-order prediction.
Ratios below unity appear toward stronger kicks and longer words.
At the same \(m\), the near-unity region extends to stronger kicks for \(b=+4\) than for \(b=+1\).
%
% 3 What meaning the pattern has :-
These trends reflect how the first-order splitting compares with the largest exact gap allowed by the Floquet zone.
At the same kick strength and word length, the larger Fourier amplitude for \(b=+1\) gives a larger predicted splitting.
Longer words narrow the Floquet zone and lower the maximum exact gap.
% 3-1 Physical picture (how strongly one word shakes the system) :-
Stronger kicks, longer words, and larger Fourier amplitudes all increase how strongly a single word shakes the system, as we quantify below.

% >> Long-time population
% 1 What the figure is about :-
We next examine whether the same Fourier amplitude also sets the population dynamics.
The red squares in Fig.~\ref{fig:fig3}(a) select parameters from the white and blue regions.
Figures~\ref{fig:fig3}(c)--(e) show the upper-band population over 30 repetitions of a Fibonacci word, obtained by multiplying the exact step operators in Eq.~\eqref{eq:step_operators}.
Their parameters lie in the near-unity region of Fig.~\ref{fig:fig3}(a), labeled as the resolved regime in Fig.~\ref{fig:fig3}(b).
% 2 What the key pattern is (build up) :-
In all three cases, the population oscillates fully between the two bands, with a well-defined period that depends on the parameters.
Each inset spectrum shows a single dominant peak, consistent with nearly sinusoidal oscillations.
% 3 What meaning the pattern has (compared to what) :-
These oscillations arise from the same resonant coupling that opens the quasienergy gap in Eq.~\eqref{eq:unfolded_quasienergy_splitting}.
For the population in Eq.~\eqref{eq:population}, the weak-kick prediction after \(r\) repetitions is
\begin{gather}\label{eq:fibonacci_rabi_probability}
    P(r\mathcal{T}_m,k_b) \simeq \sin^{2}\!\left[\frac{\pi r\mathcal{T}_m}{TN_b} \right],
    \quad
    N_b\simeq\frac{2\pi}{T\varDelta{E}_b(m)},
\end{gather}
where \(N_b\) is the oscillation period in time steps.
The blue arrows in Figs.~\ref{fig:fig3}(c)--(e) mark these predicted periods, which agree with the observed ones.
Thus, in this regime, the Rabi frequency \(2\pi/(TN_b)\) equals the exact quasienergy gap, and both equal the unfolded splitting in Eq.~\eqref{eq:unfolded_quasienergy_splitting}.
As in any resonantly coupled two-level system, the Rabi frequency equals the splitting of the coupled levels; here, this splitting is set by the temporal Fourier amplitude of an aperiodic word.
Away from \(k_b\), detuning changes both the quasienergy separation and the oscillation amplitude.
Appendix~\ref{app:detuning} derives their leading dependence and the resonance width.

% > Dash-dotted curves
% 1 What the figure is about :-
The black dash-dotted curves in Figs.~\ref{fig:fig3}(a) and (b) separate the two regimes and have both a spectral and a dynamical meaning.
They satisfy
\begin{gather}\label{eq:gap_rabi_boundary}
    \lambda|\mathcal{S}(\omega(k_b),\mathcal{T}_m)|
    =\frac{\pi}{2}.
\end{gather}
% 2-1 Physical meaning (Bloch-sphere picture of one word) :-
In the first-order resonant description, one word rotates the Bloch vector by \(2\lambda|\mathcal{S}|\) about an equatorial axis.
The curves therefore correspond to a predicted half-turn from the lower-band to the upper-band state.
% 3 Spectral meaning :-
This half-turn gives an unfolded splitting of \(\pi/\mathcal{T}_m\), which also sets the upper bound of the exact gap measured along the shorter separation of the quasienergy circle.
On the stronger-kick side, the unfolded splitting exceeds this bound.
Thus, quasienergy folding alone is sufficient to make the gap ratio \(\varDelta{E}^{\rm num}_{b}/\varDelta{E}_{b}\) smaller than unity:
\begin{gather}
	\frac{\varDelta{E}^{\rm num}_{b}}{\varDelta{E}_{b}} \leq \frac{\pi}{2\lambda|\mathcal{S}(\omega(k_b),\mathcal{T}_m)|} \leq 1.
\end{gather}
% 4 Dynamical meaning :-
The corresponding Rabi period is \(N_b=2L_m\) [Eq.~\eqref{eq:fibonacci_rabi_probability}], so sampling once per word gives two samples per predicted period along the curves.
On the weaker-kick side, \(N_b > 2L_m\) gives more than two samples per period, defining the \emph{resolved regime}.
On the stronger-kick side, \(N_b<2L_m\) gives fewer than two, defining the \emph{undersampled regime}.

% >> Single-word population
% 1 What the figure is about :-
Figures~\ref{fig:fig3}(f)--(h) resolve the population within a single Fibonacci word.
Their parameters lie in the undersampled regime of Fig.~\ref{fig:fig3}(a).
% 2 What the key pattern is :-
In all three cases, the population oscillates several times within one word, but the character of the oscillation differs.
In Fig.~\ref{fig:fig3}(h), the population oscillates nearly sinusoidally, and the inset spectrum shows a single dominant peak with weak satellites.
In Figs.~\ref{fig:fig3}(f) and (g), at stronger kicks, the oscillations are irregular, and the inset spectra show many closely spaced peaks.
% 3 What meaning the pattern has (compared to what) :-
Figure~\ref{fig:fig3}(h) shows that a gap ratio below unity can coexist with nearly sinusoidal oscillations.
% Since the first \(L_{15}\) steps of the nonrepeating drive coincide with \(w_{15}\), this trace also shows its first \(L_{15}\) steps.
The weak satellites indicate deviations from a single-frequency description, which become more pronounced in Figs.~\ref{fig:fig3}(f) and (g).
Higher-order processes involving successive kicks can modify the first-order resonant dynamics as the kick strength increases.
The undersampled regime therefore includes both nearly sinusoidal oscillations and more complex dynamics.

% Bragg peaks and convergence :-
Finally, we examine how the approximant spectra relate to interband population dynamics under the nonrepeating drive as \(m\to\infty\).
The Fibonacci harmonics approach the Bragg peaks of the kick sequence, the temporal analogues of diffraction peaks in a quasicrystal.
At a Bragg peak, the magnitude of the Fourier sum, Eq.~\eqref{eq:bragg_amplitude}, and the word duration \(\mathcal{T}_m=L_mT\) both grow linearly with \(L_m\).
For fixed \(b\neq0\), their ratio converges, and Eq.~\eqref{eq:unfolded_quasienergy_splitting} gives the finite limit
\begin{gather}\label{eq:gap_limit}
    \varDelta{E}_b(m)
    \to\frac{2\lambda}{T}
    \frac{|\sin(\pi b\phi^{-2})|}{\pi|b|}.
\end{gather}
% Order of limits and the nonrepeating drive :-
The exact gap is bounded by \(\pi/\mathcal{T}_m\) and therefore vanishes as \(m\to\infty\).
The gap ratio thus depends on the order of limits: it tends to one as \(\lambda\to0\) at fixed \(m\), but to zero as \(m\to\infty\) at fixed \(\lambda>0\).
%
% 2 What the nonrepeating drive shows :-
To connect this spectral limit to the nonrepeating dynamics, we use the convergence of the normalized Fourier amplitude along all integer times.
Averaging the interaction-picture evolution with this limit gives
\begin{gather}\label{eq:nonrepeating_rabi}
    P(t_n,k)\simeq
    \sin^{2}\!\left[
        \,\lambda\,\frac{|\sin(\pi b\phi^{-2})|}{\pi|b|T}\,t_{n}\,
    \right]
\end{gather}
for fixed \(b\neq0\) and an accessible momentum satisfying \(\omega(k)T\equiv2\pi b\phi^{-2}\pmod{2\pi}\).
Although the response under the nonrepeating drive is generically not quasiperiodic~\cite{luck1988on}, Eq.~\eqref{eq:nonrepeating_rabi} holds uniformly over any fixed number of Rabi periods as \(\lambda\to0\).
% 3 What meaning it has :-
The limiting unfolded splitting in Eq.~\eqref{eq:gap_limit} thus sets the angular frequency of the interband oscillations under the nonrepeating drive.
A detailed derivation is given in Appendix~\ref{app:bragg}.

% SECTION : HALF-INTEGER
\section{Half-integer precession and logarithmic accumulation}
\label{sec:half_integer}
 
% > Section overview
% 1 What the section is about :-
%The first-order analysis of Sec.~\ref{sec:quasienergy_resonances} holds only over a fixed number of Rabi periods, so it cannot address the second question, which concerns much longer times.
In this section, we compare the interband dynamics under repeated Fibonacci words and the nonrepeating drive at special momenta.
We first derive the exact population and classify the responses under repeated words.
We then compare the times needed to first reach a fixed upper-band population under the two drives.
Finally, we examine the effect of detuning from these momenta.

\begin{figure}[t]
    \centering
    \includegraphics[width=\columnwidth]{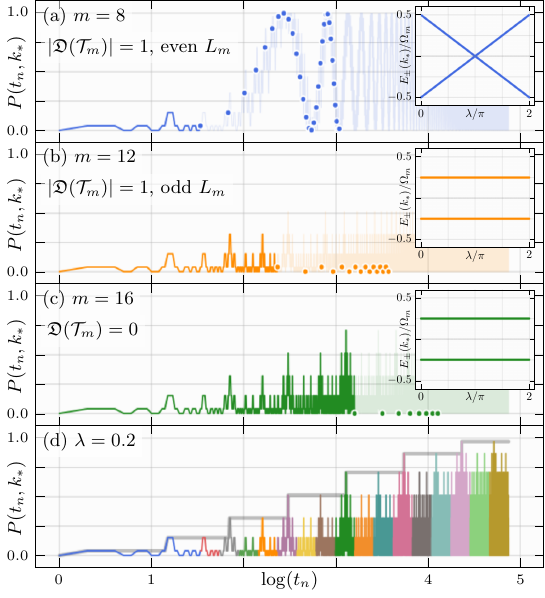}
    \caption{
    \textbf{Interband dynamics at the special momenta \(k_*\).}
    (a)--(c) Exact upper-band populations under repeated Fibonacci words with \(m=8,12,16\), respectively.
    Dots mark the population at the end of each word, which oscillates in (a), returns to zero every two words in (b), and vanishes after every word in (c).
    Solid and faint curves show the first word and later repetitions.
    Insets show \(E_\pm(k_*)/\Omega_m\) versus \(\lambda\) over \(0\leq\lambda\leq2\pi\), the period of the kick operator.
    (d) Exact upper-band population under the nonrepeating drive.
    Colored curves show kick-resolved populations; the black staircase shows their running maximum.
    We set \(J=T=1\).
    All population traces start from the lower free band at \(\lambda=0.2\), with \(k_*\) satisfying Eq.~\eqref{eq:half_integer_condition}.
    Here, \(\Omega_m=2\pi/\mathcal T_m\), and time \(t_n/T\) is shown on a base-10 logarithmic scale.
    }
    \label{fig:fig4}
\end{figure}

% > Half-integer precession and exact dynamics
% 1 Condition :-
We consider momenta \(k=k_*\) at which one step of free evolution changes the relative phase between the two bands by an odd multiple of \(\pi\):
\begin{gather}\label{eq:half_integer_condition}
    |\epsilon(k_*)|T = \left(s+\frac{1}{2}\right)\pi, \quad s=0,1,2,\ldots.
\end{gather}
The allowed values of \(s\) satisfy \((s+1/2)\pi\leq2JT\).
For \(J=T=1\), only \(s=0\) is allowed.
We refer to Eq.~\eqref{eq:half_integer_condition} as the half-integer precession condition.
At these momenta, free evolution over one step is \(F(k_*)=i(-1)^{s}\tau_z\).
%\(F(k_*)=i(-1)^{s}\tau_z\).

% 2 Commuting kick operators :-
Because \(\tau_z\) anticommutes with \(\tau_x\), one step of free evolution reverses the sign of the interband coupling, \(F(k_*)\tau_xF^{-1}(k_*)=-\tau_x\).
Since all kicks act through \(\tau_x\), kicks separated by an even number of steps add, whereas kicks separated by an odd number of steps subtract.
The net effect after \(n\) steps is therefore set by the kick-count imbalance, which is defined as
\begin{gather}\label{eq:alternating_sum}
    \mathfrak{D}(t_{n}) := \sum_{j=1}^{n}(-1)^{n-j}f_{j} = (-1)^n\mathcal{S}(\omega_\pi,t_n),
\end{gather}
where \(\omega_\pi=\pi/T\).
Since \(\omega(k_*)T=(2s+1)\pi\), the phase factor \(\exp[-i\omega(k_*)t_j]\) equals \((-1)^j\) at every kick, as it does at \(\omega_\pi\).
Up to a sign, the imbalance is therefore the temporal Fourier amplitude at the transition frequency \(\omega(k_*)\), and it is the same for all \(s\).
% By contrast, for integer precession all kicks add with the same sign, so their order is irrelevant (Appendix~\ref{app:exact}).
% Apart from that case, the half-integer condition is the only one under which all kicks commute in the interaction picture, which makes the exact solution below possible.
% In the language of dynamical decoupling, each free step at \(k_*\) acts as a \(\pi\) rotation about \(\tau_z\) between kicks.
% The imbalance \(\mathfrak{D}(t_n)\) is then the sum of the kick signs in the interaction picture with respect to free evolution~\cite{viola1999dynamical,souza2012robust}.

% 3 Exact solution :-
For integer precession, by contrast, all kicks add with the same sign, and their order is irrelevant (Appendix~\ref{app:exact}).
Apart from that case, the half-integer condition is the only one under which all kicks commute in the interaction picture.
The exact propagator up to time \(t_{n}\) is then
\begin{gather}\label{eq:half_integer_propagator}
    \mathcal{U}(t_{n},k_{*}) = F^{n}(k_{*}) \exp\!\left[ -i\lambda(-1)^n\mathfrak{D}(t_{n})\tau_{x} \right].
\end{gather}
Since \(F^n(k_*)\) only changes the phases of the free-band states, substituting Eq.~\eqref{eq:half_integer_propagator} into Eq.~\eqref{eq:population} gives
\begin{gather}\label{eq:half_integer_exact_dynamics}
    P(t_n,k_*) = \sin^2\!\left[\lambda\mathfrak{D}(t_n)\right].
\end{gather}
The kick-resolved dynamics at \(k_*\) is thus determined exactly by the integer \(\mathfrak{D}(t_{n})\).
%
% 4 Relation to prior work :-
Commuting kicks also make quasiperiodically kicked two-level systems exactly solvable~\cite{feudel1995correlation,mestel2002renormalization,mestel2004renormalization}.
There, the kick signs are imposed by the modulation, and the autocorrelation is averaged over the starting phase of the sequence.
Here, all kicks have the same sign, free evolution at \(k_*\) supplies the alternating signs, and switching on the drive at \(t=0\) selects a single trajectory.
 
% > Three classes of Fibonacci words
% 1 What sets the response :-
Under repeated words, the response at \(k_*\) depends on two properties of a word: the parity of its length \(L_m\) and its endpoint imbalance \(\mathfrak{D}(\mathcal{T}_m)\).
The concatenation rule for Fibonacci words defined in Sec.~\ref{sec:model} gives
\begin{gather}\label{eq:imbalance_cyclic}
	\mathfrak{D}(\mathcal{T}_{m+3})=-\mathfrak{D}(\mathcal {T}_m),
\end{gather}
while \(L_{m+3}\) and \(L_m\) have the same parity, as shown in Appendix~\ref{app:half_integer}.
% 2 Classification :-
Therefore, Fibonacci words with \(m\geq2\) fall into three classes according to \(m\bmod3\):
\begin{gather}\label{eq:word_classes}
\begin{aligned}
    \text{Class~I} :&   && m\equiv2, && L_m\ \text{even}, && |\mathfrak{D}(\mathcal T_m)|=1,\\
    \text{Class~II} :&  && m\equiv0, && L_m\ \text{odd},  && |\mathfrak{D}(\mathcal T_m)|=1,\\
    \text{Class~III} :& && m\equiv1, && L_m\ \text{odd},  && \phantom{|}\mathfrak{D}(\mathcal T_m)\phantom{|}=0.
\end{aligned}
\end{gather}
The words \(m=8,12,16\) in Figs.~\ref{fig:fig4}(a)--(c) represent Classes~I--III, respectively.
% 3 Relation to folded crossings :-
For even \(L_m\), the one-word free phase \(|\epsilon(k_*)|\mathcal T_m\) is an integer multiple of \(\pi\), giving a folded free-band crossing (Sec.~\ref{sec:quasienergy_resonances}).
For odd \(L_m\), it is a half-integer multiple of \(\pi\), so the two free quasienergies remain distinct.
Only Class~I words therefore place \(k_*\) at a folded crossing.
 
% > Repeated words with nonzero endpoint imbalance: (a) and (b)
% 1 What the figure is about :-
Figures~\ref{fig:fig4}(a) and (b) compare the upper-band populations under repeated Fibonacci words with \(m=8\) and \(12\), respectively.
Both words have \(|\mathfrak{D}(\mathcal T_m)|=1\) but lengths of opposite parity.
% 2 What the key pattern is :-
The endpoint population, marked by dots, oscillates across repetitions in Fig.~\ref{fig:fig4}(a) but returns to zero every two words in Fig.~\ref{fig:fig4}(b).
% 3 What meaning the pattern has (compared to what) :-
Successive repetitions are separated by \(L_m\) steps, so the alternating signs in Eq.~\eqref{eq:alternating_sum} apply to whole words.
For even \(L_m\), the contributions of successive repetitions add, whereas for odd \(L_m\), they cancel in pairs:
\begin{gather}\label{eq:periodic_closure_probability}
    P(r\mathcal T_m,k_*)=
    \begin{cases}
       \sin^{2}(r\lambda\mathfrak{D}(\mathcal T_m)),
           & \!\!\!\text{even}\ L_m,\\[8pt]
       \displaystyle\sin^{2}\frac{r\pi}{2}
       \sin^{2}(\lambda\mathfrak{D}(\mathcal T_m)),
           & \!\!\!\text{odd}\ L_m.
    \end{cases}
\end{gather}
% 4 Accumulation under repeated words :-
For Class~I, the endpoint imbalance after \(r\) repetitions is \(r\mathfrak{D}(\mathcal{T}_m)\).
Reaching a fixed upper-band population \(0 < \mathcal{P} < 1\) then requires \(r\simeq\arcsin\sqrt{\mathcal{P}}/\lambda\) repetitions, so the time to first reach it is
\begin{gather}\label{eq:periodic_waiting_time}
    \tau_{\texttt{I}}\simeq\frac{\arcsin\!\sqrt{\mathcal{P}}}{\lambda}\mathcal{T}_{m}
\end{gather}
as \(\lambda\to0\) with \(m\) fixed.
For Class~II, two repetitions cancel the imbalance, \(\mathfrak{D}(2\mathcal{T}_m)=0\), so the population never grows.
 
% 5 Quasienergy insets of (a) and (b) :-
The insets in Figs.~\ref{fig:fig4}(a) and (b) show the normalized quasienergies of the same two words as \(\lambda\) varies.
For \(m=8\), they split from the free crossing at the zone boundary; for \(m=12\), they remain fixed at \(\pm1/4\).
The exact normalized quasienergies are
\begin{gather}\label{eq:half_integer_quasienergies}
    \frac{E_{\pm}(k_*)}{\Omega_m}
    =
    \begin{cases}
        \displaystyle\frac{(2s+1)L_m}{4}
        \pm\frac{\lambda}{2\pi},
            & \!\!\!\text{even}\ L_m,\\[10pt]
        \displaystyle\pm\frac{1}{4},
            & \!\!\!\text{odd}\ L_m.
    \end{cases}
\end{gather}
The even-length splitting \(2\lambda/\mathcal{T}_m\) equals the unfolded splitting in Eq.~\eqref{eq:unfolded_quasienergy_splitting} with \(|\mathcal S(\omega_\pi,\mathcal{T}_m)|=|\mathfrak{D}(\mathcal{T}_m)|=1\), so the first-order result is exact at \(k_*\).
Although \(\omega_\pi\) does not correspond to a Bragg peak and the Fourier amplitude there grows only sublinearly [Eq.~\eqref{eq:bragg_amplitude}], this amplitude determines the dynamics exactly because the kicks commute in the interaction picture.
%
% 6 Odd-length word and six-cycle :-
For the odd-length word, the quasienergies do not depend on \(\lambda\), yet the endpoint population in Fig.~\ref{fig:fig4}(b) does.
The quasienergies fix only the word-boundary oscillation frequency; its oscillation amplitude is set by the Floquet eigenstates.
For odd \(L_m\), the one-word propagator is a \(\pi\) rotation about an axis tilted from \(\tau_z\) by \(\lambda|\mathfrak{D}(\mathcal{T}_m)|\), and this tilt gives the amplitude \(\sin^2[\lambda\mathfrak{D}(\mathcal{T}_m)]\).
 
% > Repeated words with zero endpoint imbalance: (c)
% 1 What the figure is about :-
Figure~\ref{fig:fig4}(c) shows the population under the repeated Class~III word with \(m=16\).
% 2 What the key pattern is :-
Transitions occur within each word, but the population returns to zero at every word end, while the quasienergies stay at \(\pm1/4\), as in Fig.~\ref{fig:fig4}(b).
% 3 What meaning the pattern has (compared to what) :-
Since \(\mathfrak{D}(\mathcal{T}_m)=0\), the kick contributions cancel within each word, the rotation axis coincides with \(\tau_z\), and the Floquet eigenstates are the free-band states themselves.
Figures~\ref{fig:fig4}(b) and (c) thus share their quasienergies but not their Floquet eigenstates.

% Cycles in approximant order :-
Across all three classes, Eq.~\eqref{eq:periodic_closure_probability} shows that the endpoint population depends only on \(|\mathfrak{D}(\mathcal{T}_m)|\) and the parity of \(L_m\).
Both are unchanged when \(m\) increases by three [Eq.~\eqref{eq:imbalance_cyclic}], giving a three-cycle:
\begin{gather}\label{eq:half_integer_population_three_cycle}
    P(r\mathcal{T}_{m+3},k_{*}) = P(r\mathcal{T}_m,k_*).
\end{gather}
The normalized quasienergies in Eq.~\eqref{eq:half_integer_quasienergies}, in contrast, also depend on the scalar phase \((-1)^{L_m/2}\), which drops out of the population, and therefore repeat only every six approximants [Eq.~\eqref{eq:half_integer_six_cycle}].
% because they also record the scalar phase \((-1)^{L_m/2}\) of free evolution over an even-length word, which drops out of the population.
This six-cycle is consistent with that of the Fibonacci trace map at fixed \(\lambda\)~\cite{iguchi1991theoryII}; a related six-cycle occurs in the perfect-kick limit \(\lambda=\pi/2\) (Appendix~\ref{app:perfect_kick}).

% 4 Bridge to (d) :-
The transitions within each word in Figs.~\ref{fig:fig4}(a)--(c) show that the endpoint imbalance does not capture the imbalance reached inside a word, just as trace-map cycles do not constrain the evolution between word ends~\cite{luck1988on}.
For Classes~II and III, the imbalance repeats every two words and every word, respectively, so the maximum imbalance reached within each word never grows.
For Class~I, the maximum is shifted by the endpoint imbalance accumulated over the preceding repetitions.
The nonrepeating drive instead keeps appending new kicks, which can carry the imbalance beyond its previous maximum.
 
% > Nonrepeating drive: panel (d)
% 1 What the figure is about :-
Figure~\ref{fig:fig4}(d) shows the population under the nonrepeating drive.
Since the first \(L_m\) steps of the drive coincide with \(w_m\) (Sec.~\ref{sec:model}), the displayed evolution passes from \(w_8\) to \(w_{24}\) without restarting.
% 2 What the key pattern is (build up) :-
The population does not grow monotonically, but larger peaks become accessible over time.
The black staircase, the running maximum of the population, rises only when a new record peak appears.
Successive record peaks are roughly equally spaced on the logarithmic time axis, so each larger peak takes progressively longer to appear.
 
% 3 What meaning the pattern has (compared to what) :-
To quantify this accumulation, we define the maximum kick-count imbalance through \(L_m\) steps,
\begin{gather}
    \mathfrak{M}_m := \max_{ n \leq L_m }|\mathfrak{D}(t_n)|.
\end{gather}
Within an appended word, the imbalance combines the contribution of that word with the imbalance left from the preceding word.
The latter changes sign at each step, so it can reinforce or reduce the new contribution.
Tracking the largest and smallest values of the signed imbalance \((-1)^n\mathfrak{D}(t_n)\) through each word yields the recurrence
\begin{gather}\label{eq:maximum_imbalance_recursion}
	\mathfrak{M}_{m+3}=\mathfrak{M}_m+1,
\end{gather}
for \(m\geq4\), as derived in Appendix~\ref{app:half_integer}.
With \(\mathfrak{M}_4=\mathfrak{M}_5=\mathfrak{M}_6=1\), this gives, for \(m\geq4\),
\begin{gather}\label{eq:fibonacci_log_envelope}
    \mathfrak{M}_{m} = \left\lfloor\frac{m-1}{3}\right\rfloor = \frac{\ln L_m}{3\ln\phi}+\mathcal{O}(1).
\end{gather}
Each increase of \(m\) by three multiplies the elapsed time by about \(\phi^3\), which explains the approximately equal spacing of the record peaks in Fig.~\ref{fig:fig4}(d).
%
% Relation to the spatial height function :-
The imbalance \(\mathfrak{D}\) is not specific to the driven problem.
In the spatial Fibonacci chain, the same signed count appears as the height function that sets the amplitudes of the band-center state, and its maximum also grows logarithmically~\cite{kohmoto1987critical,mace2017critical}.
For Fibonacci kicks at generic parameters, numerical data on polarization fluctuations were found compatible with logarithmic growth~\cite{luck1988on}.
Here, the same count controls a dynamical observable, the interband population, exactly.
 
% 4 Connection to population :-
Substituting Eq.~\eqref{eq:fibonacci_log_envelope} into Eq.~\eqref{eq:half_integer_exact_dynamics} connects this logarithmic accumulation to the population peaks.
As long as \(\lambda\mathfrak{M}_m\leq\pi/2\), the maximum population through \(L_m\) steps is exactly \(\sin^{2}(\lambda\mathfrak{M}_m)\).
For the displayed kick strength \(\lambda=0.2\), this condition holds through \(m=24\), where \(\mathfrak{M}_{24}=7\).
The running population maximum then reaches \(0.97\), consistent with the black staircase.
Beyond \(\lambda\mathfrak M_m=\pi/2\), the running maximum approaches unity if \(\lambda/\pi\) is irrational, as for the value \(\lambda = 0.2\).
For rational \(\lambda/\pi\), the largest accessible population can be smaller than one.
 
% 5 Comparison with repeated words :-
% 5-1 What sets the waiting time :-
We now compare how long each drive takes to first reach a fixed upper-band population \(0<\mathcal{P}<1\).
Let \(\tau_{\rm F}\) denote this waiting time under the nonrepeating drive.
The population in Eq.~\eqref{eq:half_integer_exact_dynamics} increases with the imbalance magnitude for \(\lambda|\mathfrak{D}(t_n)|\leq\pi/2\).
Then, reaching \(\mathcal{P}\) requires
\begin{gather}
    |\mathfrak{D}(t_n)|
    \geq\frac{\arcsin\sqrt{\mathcal{P}}}{\lambda}.
\end{gather}
Therefore, its smallest required imbalance magnitude is
\begin{gather}\label{eq:smalled_required_imbalance}
    M_{*}:=\left\lceil
        \frac{\arcsin\sqrt{\mathcal{P}}}{\lambda}
    \right\rceil,
\end{gather}
where \(\lceil x\rceil\) is the smallest integer greater than or equal to \(x\).
For sufficiently weak kicks, \(\lambda M_{*}\leq\pi/2\), so this threshold lies within the increasing range.
The imbalance magnitude changes by at most one per step and cannot skip \(M_{*}\).
Thus, the population first reaches or exceeds \(\mathcal{P}\) when the imbalance magnitude first reaches \(M_{*}\).
% 5-2 Role of logarithmic accumulation :-
Under the nonrepeating drive, the maximum imbalance increases by one every three word orders [Eq.~\eqref{eq:maximum_imbalance_recursion}], while the word duration grows by a factor of about \(\phi^{3}\).
% Binet formula \(L_m=\frac{\phi^{m+1}-(-\phi^{-1})^{m+1}}{\sqrt{5}}\)
% \(L_m\sim\frac{\phi^{m+1}}{\sqrt{5}}\quad(m\to\infty)\)
% 5-3 Quantitative statement :-
Equation~\eqref{eq:fibonacci_log_envelope} gives the maximum imbalance reached through \(L_m\) steps.
Between \(L_m\) and \(L_{m+1}\) steps, the running maximum lies between \(\mathfrak{M}_{m}\) and \(\mathfrak{M}_{m+1}\).
These bounds differ by at most one, and successive word lengths differ by a bounded factor.
The logarithmic estimate in Eq.~\eqref{eq:fibonacci_log_envelope} therefore extends to every integer time:
\begin{gather}\label{eq:fibonacci_log_envelop_generalized}
    \max_{j\leq n}|\mathfrak{D}(t_j)| = \frac{\ln n}{3\ln\phi}+\mathcal{O}(1).
\end{gather}
At \(n=\tau_{\rm F}/T\), the running maximum equals \(M_{*}\).
Using these values in Eq.~\eqref{eq:fibonacci_log_envelop_generalized} and \(M_{*}\) from Eq.~\eqref{eq:smalled_required_imbalance}, we obtain
\begin{align}
    \ln\frac{\tau_{\rm F}}{T}
    &= 3M_{*}\ln\phi+\mathcal{O}(1) \notag\\[3pt]
    &= \frac{3\ln\phi\,\arcsin\!\sqrt{\mathcal{P}}}{\lambda}+\mathcal{O}(1) \label{eq:fibonacci_waiting_time}
\end{align}
as \(\lambda\to0\).
% 5-4 Meaning :-
The nonrepeating drive therefore takes a time exponential in \(1/\lambda\) to reach the target population.
In contrast, repeating a fixed Class~I word makes the endpoint imbalance magnitude increase by one per word, giving a waiting time proportional to \(1/\lambda\) [Eq.~\eqref{eq:periodic_waiting_time}].

% Detuning :-
We finally examine how detuning from \(k_*\) affects the population.
Writing \(|\epsilon(k)|T=(s+1/2)\pi+\delta\), the detuning adds a relative phase \(2\delta\) per free step.
These phase differences accumulate over time.
At any kick strength, the resulting population error is bounded by \(2n|\delta|\); see Appendix~\ref{app:detuning_half_integer} for detailed discussions.
A sufficient condition for the population to remain close to the resonant trajectory through the waiting time \(\tau_{\rm F}\) is therefore
\begin{gather}\label{eq:detuning_requirement}
    |\delta|\ll\frac{T}{\tau_{\rm F}} = \exp\!\left[ -\frac{3\ln\phi\,\arcsin\!\sqrt{\mathcal{P}}}{\lambda} + \mathcal{O}(1) \right].
\end{gather}
This sufficient window narrows inversely with the observation time.
For repetition of a fixed Class~I word, the weak-kick resonance half-width is \(|\delta|\simeq\lambda/L_m\), which is likewise of the order of \(T/\tau_{\texttt{I}}\).

% SECTION : CONCLUSIONS
\section{Conclusions}
\label{sec:discussions}
 
% Main message :-
We asked what the spectra of periodic approximants reveal about a lattice driven by Fibonacci-ordered kicks.
% Our answer has two parts.
%
% Claim 1 :-
For weak kicks, the approximant spectrum captures the resonant dynamics.
At each resonance, a single temporal Fourier amplitude of the kick sequence sets both the quasienergy splitting and the Rabi frequency.
Where this amplitude grows in proportion to time, the splitting converges as the approximant grows and sets the Rabi frequency of the nonrepeating drive, even though the exact gap of the approximant vanishes.
%
% Claim 2 :-
The approximant spectrum, however, misses how population accumulates within the words.
At special momenta, the upper-band population is fixed exactly by a signed count of kicks on even and odd steps.
Under the nonrepeating drive, the maximum of this count grows only logarithmically in time.
Reaching a fixed population therefore takes a time exponential in \(1/\lambda\), compared with linear in \(1/\lambda\) for a repeated even-length word.
The same signed count describes the height function of the band-center state of the spatial Fibonacci chain; here, it determines an upper-band population and a waiting time.
 
% Experimental prospects :-
Our findings could be tested with ultracold atoms in optical lattices, where pulsed lattices have been implemented~\cite{shimasaki2024anomalous}.
Fibonacci-ordered pulse sequences have also been realized in trapped ions~\cite{dumitrescu2022dynamical}.
A lattice realization requires a pulsed secondary lattice with twice the period of the primary lattice~\cite{sebby2006lattice}, and band mapping to resolve the upper-band population at each momentum~\cite{greiner2001exploring}.
The logarithmic accumulation is the most demanding to observe, because it requires long coherence times and a detuning from the special momenta well below \(T/\tau_{\rm F}\) [Eq.~\eqref{eq:detuning_requirement}].
 
% Outlook :-
Interactions couple different momentum sectors, so it remains open whether the exact cancellation at the special momenta survives in interacting systems.
For weak kicks, transition amplitudes still involve the same Fourier amplitude, so Fibonacci-ordered kicks might suppress selected breakup channels of repulsively bound pairs~\cite{winkler2006repulsively}.
Other aperiodic kick sequences, such as Thue--Morse ordering, have different Fourier amplitudes and are also worth examining.

\begin{acknowledgements}
This research was supported by an appointment to the JRG Program at the APCTP through the Science and Technology Promotion Fund and Lottery Fund of the Korean Government, the Korean Local Governments (Gyeongsangbuk-do Province and Pohang City), and the National Research Foundation of Korea (NRF) funded by the Korean government (Ministry of Science and ICT, MSIT) (No. RS-2026-25499525).
\end{acknowledgements}

\appendix

% APPENDIX : SECTION A
\section{Model symmetries}
\label{app:symmetries}
 
Changing \(\lambda\) to \(-\lambda\) in the model [Eq.~\eqref{eq:model}] corresponds to translating the staggered potential by one lattice site.
With the \(\lambda\) dependence made explicit, the kick operators in Eq.~\eqref{eq:step_operators} satisfy
\begin{gather}
\begin{aligned}
    K_n(-\lambda) &= \tau_z\,K_n(\lambda)\,\tau_z,\\[3pt]
    K_n(\lambda+\pi) &= (-1)^{f_n}K_n(\lambda).
\end{aligned}
\end{gather}
Since \(F(k)\) commutes with \(\tau_z\), the first relation conjugates the full propagator by \(\tau_z\) and preserves the quasienergies.
It also preserves the population in Eq.~\eqref{eq:population}, because the initial and final free-band states are eigenstates of \(\tau_z\).
The second relation multiplies each active kick operator by \(-1\) under \(\lambda\to\lambda+\pi\).
The full propagator therefore acquires an overall sign, which cancels in the absolute square defining the population.
A Fibonacci word contains \(L_{m-2}\) active kicks.
When \(L_{m-2}\) is odd, the shift \(\lambda\to\lambda+\pi\) multiplies the one-word propagator \(\mathcal{U}(\mathcal{T}_{m},k)\) by \(-1\).
Both quasienergies therefore shift by \(\Omega_m/2\), modulo \(\Omega_m\).
Their shorter separation on the quasienergy circle is unchanged, so the exact gap is preserved.
Increasing \(\lambda\) by \(2\pi\) restores each kick operator and hence the full quasienergy spectrum.
The populations and quasienergy gaps are therefore \(\pi\)-periodic in \(\lambda\) and invariant under \(\lambda\to-\lambda\).
Together, these symmetries also imply invariance under \(\lambda\to\pi-\lambda\), justifying the restriction to \(0\leq\lambda\leq\pi/2\).
 
% APPENDIX : SECTION B
\section{Dynamics at integer precession}
\label{app:exact}
 
Let \(p=sL_m\) in the crossing condition of Eq.~\eqref{eq:folded_crossing_condition}, where \(s\) is a nonnegative integer.
Since \(\mathcal T_m=L_mT\), these momenta satisfy \(|\epsilon(k)|T=s\pi\).
The allowed momenta are
\begin{gather}
    k=\pm2\arccos\left(\frac{s\pi}{2JT}\right),
    \quad 0\leq s\pi\leq2JT.
\end{gather}
All these crossings satisfy the \(b=0\) selection condition in Eq.~\eqref{eq:temporal_bragg_condition}, including the original free-band touching at \(k=\pm\pi\) for \(p=s=0\).
For \(2JT<\pi\), only the band-touching case \(s=0\) is allowed.
At these momenta, the one-step free evolution is \(F(k)=(-1)^{s}\mathbb{I}\).
The free evolution between kicks contributes only a common phase, so all active kicks accumulate about the \(\tau_{x}\) axis.
Since the word contains \(L_{m-2}\) active kicks, the exact one-word propagator is
\begin{gather}\label{eq:band_touching_propagator}
    \mathcal U(\mathcal{T}_m,k) =(-1)^{sL_m} \exp(-i\lambda L_{m-2}\tau_x),
\end{gather}
which depends only on the number of active kicks and the parity of \(sL_{m}\), not on their ordering.
Substituting Eq.~\eqref{eq:band_touching_propagator} into Eq.~\eqref{eq:diagonalization} gives the quasienergies
\begin{gather}
    E_{\pm}(k) =\frac{s\pi}{T} \pm\frac{\lambda}{T}\frac{L_{m-2}}{L_m} \pmod{\Omega_{m}}.
\end{gather}
The first term comes from the common free-evolution phase and locates the unperturbed crossing at the zone center when \(sL_m\) is even and at the zone boundary when \(sL_m\) is odd.
The second term gives the opposite quasienergy shifts of the two \(\tau_x\) eigenstates.
The exact splitting before quasienergy folding is therefore
\begin{gather}\label{eq:quasienergy_splitting}
    \varDelta{E}_{0}(m) = \frac{2|\lambda|}{\mathcal{T}_m}L_{m-2}.
\end{gather}
 
Repeating the word \(r\) times multiplies the accumulated kick angle by \(r\).
The common phase drops out of the transition probability.
Hence, substituting Eq.~\eqref{eq:band_touching_propagator} into Eq.~\eqref{eq:population} gives the endpoint population
\begin{gather}\label{eq:exact_stroboscopic_population}
    P(r\mathcal{T}_m,k) = \sin^{2}(r\lambda L_{m-2}).
\end{gather}
This expression is exact for arbitrary \(\lambda\) and, at word boundaries, can equivalently be written in the resonant Rabi form with the exact quasienergy gap as its angular frequency.
Within each word, free evolution preserves the band populations, so population changes occur only at active kicks, producing a kick-resolved staircase.
 
% APPENDIX : SECTION C
\section{Fourier amplitude at Bragg peaks}
\label{app:bragg}

% Fibonacci sequence and Fourier amplitude :-
The infinite Fibonacci word is a Sturmian sequence, which can be generated by an irrational rotation~\cite{lothaire2002algebraic,baake2013aperiodic}:
\begin{gather}
    x_j=\{j\phi^{-2}+\rho\},
    \quad
    f_j=
    \begin{cases}
        1, & x_j\in \mathcal{I},\\[3pt]
        0, & x_j\notin \mathcal{I}.
    \end{cases}
\end{gather}
Here, \(\{\cdot\}\) removes the integer part, keeping \(x_j\) in \([0,1)\).
The fixed offset \(\rho\) sets the starting position of the rotation.
The interval \(\mathcal{I}\subset[0,1)\) has length \(1/\phi^{2}\) and selects the steps at which a kick is applied.
We use \(\mathcal{S}(\omega,t_n)\) for the Fourier sum in Eq.~\eqref{eq:fibonacci_structure_factor} over the first \(n\) time steps.

We define the Bragg peaks of the nonrepeating drive by \(\omega_bT \equiv 2\pi b/\phi^{2}\) modulo \(2\pi\), the \(m\to\infty\) limit of the Fibonacci harmonics \(b\nu_m\).
At these frequencies, the phase factor becomes
\begin{gather}
    \exp\left(-i\omega_b t_j\right) = \exp\left(2\pi ib\rho\right)\exp\left(-2\pi ibx_j\right).
\end{gather}
Since the points \(x_j\) are equidistributed on the unit interval, the normalized Fourier amplitude converges to
\begin{align}
    C_{b} &:= \lim_{n\to\infty}\!\!\frac{\mathcal{S}(\omega_{b},t_n)}{n} \notag\\[3pt]
    &\phantom{:}= \exp\left(2\pi ib\rho\right)\!\int_{\mathcal{I}}\!\exp\left(-2\pi ibx\right)dx.
\end{align}
At other frequencies, the normalized Fourier amplitude tends to zero (see also Ref.~\cite{luck1988on}).
To show this, approximate the interval indicator by trigonometric polynomials.
After multiplication by \(\exp[-i\omega t_j]\), each term has zero long-time average.
Combining the two cases and taking the magnitude gives
\begin{gather}\label{eq:bragg_amplitude}
\raisetag{2.7\baselineskip}
    \lim_{n\to\infty}\!\!\frac{|\mathcal{S}(\omega,t_n)|}{n}
    =
    \begin{cases}
        \dfrac{|\sin(\pi b\phi^{-2})|}{\pi|b|},
            & \omega T \equiv 2\pi b/\phi^{2}, \\[8pt]
        1/\phi^{2},
            & \omega T \equiv 0,\\[6pt]
        0\vphantom{\dfrac{0}{0}},
            & \text{otherwise}.
    \end{cases}
\end{gather}
All congruences are modulo \(2\pi\), and \(b\in\mathbb{Z}\setminus{\{0\}}\).
For \(b=+1\) and \(+4\), Eq.~\eqref{eq:bragg_amplitude} gives \(0.297\) and \(0.079\), respectively.

We now use Eq.~\eqref{eq:bragg_amplitude} to obtain the approximant limit in Eq.~\eqref{eq:gap_limit}.
We compare the Fourier sums at \(\omega(k_b)\) and \(\omega_b\) over the same word.
Frequencies differing by \(2\pi/T\) give the same Fourier sum, so we choose equivalent frequencies with the smallest difference.
For fixed \(b\), Binet's formula gives \(L_{m-2}/L_m=\phi^{-2}+\mathcal O(L_m^{-2})\).
Choosing matching frequency representatives modulo \(2\pi/T\), we obtain
\begin{gather}
	T\left|\omega(k_b)-\omega_b\right| = \mathcal{O}(L_m^{-2}).
\end{gather}
Applying \(|\exp[ix]-\exp[iy]|\!\leq\!|x-y|\) to each term, we get %|sin(x)| \leq |x| 
\begin{align}
    \frac{\left|\mathcal{S}(\omega(k_b),\mathcal{T}_m) - \mathcal{S}(\omega_b,\mathcal{T}_m)\right|}{L_m}
    &\leq
    \frac{L_m+1}{2}T \left|\omega(k_b)-\omega_b\right| \notag\\[3pt]
    &= \mathcal{O}(L_m^{-1}).
\end{align}
The normalized Fourier amplitudes therefore have the same limit.
Multiplying this limit by \(2\lambda/T\) gives the first-order splitting limit in Eq.~\eqref{eq:gap_limit}.

% Nonrepeating drive :-
Next, we derive Eq.~\eqref{eq:nonrepeating_rabi} for the nonrepeating drive in the weak-kick limit.
Fix a nonzero harmonic \(b\) and an accessible momentum \(k\) satisfying \(\omega(k)\equiv\omega_b\) modulo \(2\pi/T\).
In the interaction picture with respect to free evolution, we write \(\mathcal{U}(t_{n},k) = F^{n}(k)W_n\).
Then the evolution \(W_{n}\) is
\begin{gather}\label{eq:Wn}
\begin{gathered}
    W_n=\exp\left(-i\lambda A_n\right)W_{n-1}, \quad W_{0}=\mathbb{I}, \\[3pt]
    A_{n} = f_{n}
    \begin{bmatrix}
        0 & \,\exp(-i\omega_b t_{n})\,\\[3pt]
        \,\exp(i\omega_b t_{n})\, & 0
    \end{bmatrix}.
\end{gathered}
\end{gather}
Here, \(A_n\) describes the kick after free evolution has been removed.
Its off-diagonal entries mix the two bands, with phases accumulated before the kick.
The time average of \(A_n\) follows from the Fourier limit \(C_b\):
\begin{gather}
    \mathcal{A}_b := \lim_{n\to\infty}\frac{1}{n}\sum_{j=1}^{n}A_j =
    \begin{bmatrix}
        0 & \,C_b\, \\
        \,C_b^{*}\, & 0
    \end{bmatrix}.
\end{gather}

We must check that averaging remains accurate over a Rabi period, which spans \(\mathcal O(1/\lambda)\) steps.
Consider \(n\leq K/\lambda\), where \(K > 0\) is fixed independently of \(\lambda\).
Define
\begin{gather}
    B_n:=\sum_{j=1}^{n}(A_j-\mathcal{A}_b).
\end{gather}
The Fourier limit gives \(\|B_n\|/n\to0\), so the accumulated deviation grows more slowly than the number of steps.
Expanding the one-step operators and applying summation by parts gives, uniformly on this interval,
\begin{gather}
    \left\|W_n-\exp(-i\lambda n\mathcal{A}_b)\right\| = \mathcal{O}\!\left(
        \lambda\max_{j\leq n}\|B_{j}\|+\lambda
    \right),
\end{gather}
where \(\|\cdot\|\) denotes the operator norm.
The first term vanishes because the accumulated deviation grows sublinearly.
The second bounds the quadratic expansion errors accumulated over \(\mathcal{O}(1/\lambda)\) steps.
Both therefore vanish as \(\lambda\to0\), establishing the averaged evolution over any fixed number of Rabi periods.
Thus, \(W_{n}\) is effectively
\begin{gather}
    W_{n} = \exp\left(-i\lambda n\mathcal{A}_b\right) + o(1),
\end{gather}
where \(o(1)\) denotes a little-o term whose operator norm vanishes as \(\lambda\to0\), uniformly for \(n\leq K/\lambda\).
Free evolution changes only the band phases.
Using \(\mathcal{A}_b^2 = |C_b|^2\mathbb{I}\), Eq.~\eqref{eq:population} then gives Eq.~\eqref{eq:nonrepeating_rabi} in the weak-kick limit.
 
% APPENDIX : SECTION D
\section{Quasienergies at half-integer precession}
\label{app:half_integer}
 
At the half-integer precession condition in Eq.~\eqref{eq:half_integer_condition}, the corresponding momenta are
\begin{gather}
    k_*=\pm2\arccos\left[\frac{(s+1/2)\pi}{2JT}\right],
\end{gather}
where \(s\) is a nonnegative integer with \((s+1/2)\pi\leq2JT\).
For even \(L_m\), the one-word propagator in Eq.~\eqref{eq:half_integer_propagator} is then
\begin{gather}\label{eq:half_integer_propagator_even}
    \mathcal{U}(\mathcal{T}_m,k_*) = (-1)^{L_m/2} \exp\left[-i\lambda\mathfrak{D}(\mathcal{T}_m)\tau_x\right].
\end{gather}
Applying Eq.~\eqref{eq:diagonalization} to this propagator yields the even-length result in Eq.~\eqref{eq:half_integer_quasienergies}.
Here, \(|\mathfrak D(\mathcal{T}_m)|=1\) for even-length Fibonacci words, so its sign only interchanges the two quasienergy branches.
The common scalar factor is \(+1\) when \(L_m\) is divisible by four, placing the free crossing at the zone center.
For other even lengths, it is \(-1\), placing the crossing at the zone boundary.
With the exact quasienergy gap \(\varDelta{E}_{*}(m)=2|\lambda|/\mathcal{T}_m\), substituting Eq.~\eqref{eq:half_integer_propagator_even} into Eq.~\eqref{eq:population} gives the Rabi form,
\begin{gather}
    P(r\mathcal{T}_m,k_{*}) = \sin^2\!\left[\frac{\pi r\mathcal{T}_m}{TN_{*}}\right],
    \quad N_{*}=\frac{\pi L_m}{|\lambda|},
\end{gather}
where \(r\) is a nonnegative integer and \(N_{*}\) is the Rabi period measured in elementary steps.
 
For odd \(L_m\), the factor \(\smash{F^{L_m}(k_*)}\) in Eq.~\eqref{eq:half_integer_propagator} is proportional to \(\tau_z\).
Conjugation by this factor changes the sign of \(\tau_x\) and reverses the net kick rotation, giving
\begin{gather}
    F^{L_m}(k_*)R_mF^{-L_m}(k_*)=R_m^{-1},
\end{gather}
where \(R_m=\exp\!\left[i\lambda\mathfrak{D}(\mathcal{T}_m)\tau_x\right]\).
By squaring the one-word propagator, we obtain the following relation:
\begin{align}
    \mathcal{U}^2(\mathcal{T}_m,k_*)
    &=\left[F^{L_m}(k_*)R_mF^{-L_m}(k_*)\right]
      F^{2L_m}(k_*)R_m \notag\\[3pt]
    &=R_m^{-1}F^{2L_m}(k_*)R_m \notag\\[3pt]
    &=-R_m^{-1}R_m=-\mathbb{I},
\end{align}
where \(\smash{F^{2L_m}(k_*)=-\mathbb{I}}\) for odd \(L_m\).
Thus any initial state returns after two words up to a common phase factor of \(-1\), consistent with the population return in Fig.~\ref{fig:fig4}(b).
%The eigenvalues of the one-word propagator are \(+i\) and \(-i\), giving the odd-length quasienergies in Eq.~\eqref{eq:half_integer_quasienergies}, independently of \(\lambda\) and the endpoint signed sum.
The corresponding quasienergies are independent of \(\lambda\) and the imbalance \(\mathfrak{D}(\mathcal{T}_m)\), giving the odd-length result in Eq.~\eqref{eq:half_integer_quasienergies}.
We also note that, for Fibonacci words, the case \(\mathfrak{D}(\mathcal{T}_m)=0\) occurs only for odd \(L_m\).
In this case, the initial state returns after each word up to a phase and the population obeys \(P(t_n+\mathcal{T}_m,k_*)=P(t_n,k_*)\).
 
For Fibonacci words, the concatenation rule \(w_{m+1}=w_mw_{m-1}\) gives the endpoint recursion
\begin{gather}
    \mathfrak{D}(\mathcal{T}_{m+1}) = \mathfrak{D}(\mathcal{T}_{m-1}) + (-1)^{L_{m-1}}\mathfrak{D}(\mathcal{T}_m).
\end{gather}
The initial words \(w_1=0\) and \(w_2=01\) give \(\mathfrak{D}(\mathcal{T}_1)=0\) and \(\mathfrak{D}(\mathcal{T}_2)=1\).
Since the parity of \(L_m\) repeats every three orders, iterating this recursion gives the repeating sequence of \(0,1,-1,0,-1,1\).
Values three orders apart have opposite signs, establishing Eq.~\eqref{eq:imbalance_cyclic}.
These values establish \(|\mathfrak{D}(\mathcal{T}_m)|=1\) for even \(L_m\) and show that \(\mathfrak{D}(\mathcal{T}_m)=0\) occurs only for odd \(L_m\), as used above.
The six-order repetition of the full propagator also follows from Eq.~\eqref{eq:half_integer_propagator}.
The length recursion \(L_{m+1}=L_m+L_{m-1}\) gives the repeating residues \(1,2,3,1,0,1\) modulo four, so \(L_{m+6}\equiv L_m\pmod4\).
Since \(F^{4}(k_*) = \mathbb{I}\), the free-propagation factor therefore repeats after six orders.
The six-cycle of the endpoint imbalance and the unchanged length parity also leave the kick factor in Eq.~\eqref{eq:half_integer_propagator} unchanged, giving
\begin{gather}\label{eq:half_integer_six_cycle}
    \mathcal{U}(\mathcal{T}_{m+6},k_*) = \mathcal{U}(\mathcal{T}_m,k_*).
\end{gather}
Consequently, the normalized quasienergies \(E_{\pm}(k_*)/\Omega_m\), understood modulo one, repeat every six approximant orders up to an interchange of the branches.

The same concatenation rule also determines the maximum kick-count imbalance within each word.
To prove Eq.~\eqref{eq:maximum_imbalance_recursion}, we track the minimum and maximum of \((-1)^{n}\mathfrak{D}(t_n)\), whose magnitude equals \(|\mathfrak{D}(t_n)|\).
For a step in the appended word \(w_{m-1}\), the concatenation gives
\begin{gather}\label{eq:imbalance_within_appended_word}
    \mathfrak{D}(\mathcal{T}_m+t_j) =\mathfrak{D}(t_j)+(-1)^j\mathfrak{D}(\mathcal{T}_m),
\end{gather}
for \(0\leq j\leq L_{m-1}\).
Multiplying by \((-1)^{L_m+j}\) gives
\begin{gather}
\begin{aligned}
    (-1)^{L_m+j}\mathfrak{D}(\mathcal{T}_m+t_j) = &\, (-1)^{L_m} \mathfrak{D}(\mathcal{T}_m) \\[3pt]
    &+ (-1)^{L_{m}+j}\mathfrak{D}(t_j).
\end{aligned}
\end{gather}
The appended word therefore contributes a shifted copy of the values in \(w_{m-1}\), reflected when \(L_m\) is odd.
 
We prove by induction that
\begin{gather}\label{eq:imbalance_extrema_three_step}
\begin{aligned}
    \min_{n\leq L_{m+3}}\!\!\!\!
    \left[(-1)^n\mathfrak{D}(t_n)\right]
    &=-1-\max_{n\leq L_m}\!
    \left[(-1)^n\mathfrak{D}(t_n)\right], \\[3pt]
    \max_{n\leq L_{m+3}}\!\!\!\!
    \left[(-1)^n\mathfrak{D}(t_n)\right]
    &=1-\min_{n\leq L_m}\!
    \left[(-1)^n\mathfrak{D}(t_n)\right],
\end{aligned}
\end{gather}
for \(m \geq 6\).
Assume Eq.~\eqref{eq:imbalance_extrema_three_step} holds at \(m-1\) and \(m\).
For the appended word, the concatenation rule gives
\begin{align}
    &\max_{j\leq L_{m+2}}\!\!\!\!
    \left[(-1)^{L_{m+3}+j}
    \mathfrak{D}(\mathcal{T}_{m+3}+t_j)\right]
    \notag\\[3pt]
    &\quad=(-1)^{L_{m+3}}\mathfrak{D}(\mathcal{T}_{m+3})
    +\max_{j\leq L_{m+2}}\!\!\!\!
    \left[(-1)^{L_{m+3}+j}\mathfrak{D}(t_j)\right]
    \notag\\[3pt]
    &\quad=-(-1)^{L_m}\mathfrak{D}(\mathcal{T}_m)
    +1-\!\min_{j\leq L_{m-1}}\!\!\!\!
    \left[(-1)^{L_m+j}\mathfrak{D}(t_j)\right]
    \notag\\[3pt]
    &\quad=1-\!\min_{j\leq L_{m-1}}\!\!\!\!
    \left[(-1)^{L_m+j}
    \mathfrak{D}(\mathcal{T}_m+t_j)\right].
\end{align}
The second equality uses Eq.~\eqref{eq:imbalance_cyclic}, the unchanged length parity, and the induction assumption at \(m-1\).
For odd \(L_m\), multiplication by \((-1)^{L_m}\) exchanges the minimum and maximum, so both relations in Eq.~\eqref{eq:imbalance_extrema_three_step} are used.
 
The maximum over \(w_{m+4}=w_{m+3}w_{m+2}\) is the larger of the maxima in its two parts.
Using the induction assumption at \(m\) for the first part gives
\begin{align}
    &\max_{n\leq L_{m+4}}\!\!\!\!
    \left[(-1)^n\mathfrak{D}(t_n)\right]
    \notag\\[3pt]
    &\quad=\max\left\{
    \begin{aligned}
        &1-\min_{n\leq L_m}\!\!
        \left[(-1)^n\mathfrak{D}(t_n)\right],\\[3pt]
        &1-\min_{j\leq L_{m-1}}\!\!\!\!
        \left[(-1)^{L_m+j}
        \mathfrak{D}(\mathcal{T}_m+t_j)\right]
    \end{aligned}
    \right\}
    \notag\\[3pt]
    &\quad=1-\min_{n\leq L_{m+1}}\!\!\!\!
    \left[(-1)^n\mathfrak{D}(t_n)\right].
\end{align}
The last equality combines the two parts of \(w_{m+1}=w_mw_{m-1}\).
The same calculation with minimum and maximum exchanged gives
\begin{gather}
    \min_{n\leq L_{m+4}}\!\!\!\!
    \left[(-1)^n\mathfrak{D}(t_n)\right]
    =-1-\!\!\max_{n\leq L_{m+1}}\!\!\!\!
    \left[(-1)^n\mathfrak{D}(t_n)\right].
\end{gather}
This proves Eq.~\eqref{eq:imbalance_extrema_three_step} at \(m+1\).
 
Since multiplication by \((-1)^n\) preserves the magnitude, Eq.~\eqref{eq:imbalance_extrema_three_step} gives the following relation:
\begin{align}
    \mathfrak{M}_{m+3}
    &=\max\left\{
    \begin{aligned}
        &1-\min_{n\leq L_m}
        \left[(-1)^n\mathfrak{D}(t_n)\right],\\[3pt]
        &1+\max_{n\leq L_m}
        \left[(-1)^n\mathfrak{D}(t_n)\right]
    \end{aligned}
    \right\}
    \notag\\[3pt]
    &=1+\max_{n\leq L_m}
    \left|\mathfrak{D}(t_n)\right|
    =1+\mathfrak{M}_m,
\end{align}
for \(m\geq6\).
The directly evaluated values
\(\mathfrak{M}_4=\mathfrak{M}_5=\mathfrak{M}_6=1\) and
\(\mathfrak{M}_7=\mathfrak{M}_8=2\) establish the remaining
cases \(m=4,5\).
Thus Eq.~\eqref{eq:maximum_imbalance_recursion} holds.

% APPENDIX : SECTION E
\section{Exact solution in the perfect-kick limit}
\label{app:perfect_kick}
 
At \(\lambda=\pi/2\), the kick operator becomes \(K_n = K^{f_n}\), with \(K=-i\tau_x\).
Each active kick completely exchanges the two free-band states, while free evolution changes only their phases.
The band populations therefore depend only on the number of active kicks, which allows an exact solution for any \(\epsilon(k)T\).
For \(m\geq3\), the Fibonacci word contains \(L_{m-2}\) active kicks, giving
\begin{gather}
    P(\mathcal T_m,k)=
    \begin{cases}
        0, & L_{m-2}\ \text{even},\\[3pt]
        1, & L_{m-2}\ \text{odd}.
    \end{cases}
\end{gather}
To obtain the quasienergies, we use the identity \(F(k)K = KF^{-1}(k)\).
Writing \(F=F(k)\) and \(\mathcal U_m=\mathcal U(\mathcal T_m,k)\), the word concatenation gives
\begin{gather}
    \mathcal{U}(\mathcal{T}_{m+1},k) = \mathcal{U}(\mathcal{T}_{m-1},k) \ \mathcal{U}(\mathcal{T}_m,k).
\end{gather}
The initial words \(w_1=0\) and \(w_2=01\) give \(\mathcal{U}(\mathcal{T}_{1},k) = F\) and \(\mathcal{U}(\mathcal{T}_{2},k) =KF^{2}(k)\).
Using \(K^{2}=-\mathbb{I}\) and \(F(k)K = KF^{-1}(k)\), successive iterations yield
\begin{gather}
\begin{aligned}
    \mathcal U(\mathcal T_3,k) &= KF(k),
    &\quad \mathcal U(\mathcal T_4,k) &= -F^{-1}(k),\\[4pt]
    \mathcal U(\mathcal T_5,k) &= -K,
    &\quad \mathcal U(\mathcal T_6,k) &= KF(k).
\end{aligned}
\end{gather}
These expressions give \(\mathcal{U}(\mathcal{T}_{m+6},k)=\mathcal{U}(\mathcal{T}_{m},k)\), so the propagators repeat every six approximant orders.
The diagonal cases occur at \(m\equiv1,4\pmod6\), precisely the words with an even number of active kicks.
For the remaining words, the propagator is off-diagonal and exchanges the two band states.
Its zero trace and unit determinant fix its eigenvalues to \(+i\) and \(-i\).
Applying Eq.~\eqref{eq:diagonalization} gives the normalized quasienergies
\begin{gather}\label{eq:perfect_kick_quasienergies}
    \frac{E_\pm(k)}{\Omega_m}=
    \begin{cases}
        \displaystyle
        \frac{L_{m-2}}{4}
        \pm\frac{\epsilon(k)T}{2\pi},
        & L_{m-2}\ \text{even},\\[10pt]
        \displaystyle
        \pm\frac14,
        & L_{m-2}\ \text{odd},
    \end{cases}
\end{gather}
For even \(L_{m-2}\), the quasienergies retain momentum dependence through \(\epsilon(k)\).
The scalar phase accumulated by the kicks shifts both normalized quasienergies by either zero or \(1/2\).
For odd \(L_{m-2}\), the normalized quasienergies are fixed at \(\pm1/4\), yielding two flat quasienergy bands.
 
% APPENDIX : SECTION F
\section{Detuning and resonance width}
\label{app:detuning}

At the crossing momentum \(k_b\) discussed in Sec.~\ref{sec:quasienergy_resonances}, the free evolution over one word is proportional to the identity.
Away from \(k_b\), the two bands acquire opposite phase shifts relative to the crossing.
We describe these shifts by the dimensionless detuning
\begin{gather}
    \delta_b(k) = \frac{\mathcal{T}_m}{2} [\omega(k_b)-\omega(k)].
\end{gather}
Near \(k_b\), we approximate \(\mathcal{S}(\omega(k),\mathcal{T}_m)\) by its value at the crossing.
This keeps the coupling fixed while retaining the momentum dependence of the detuning.
In the free-band basis \((\ket{+,k},\ket{-,k})\), the effective Floquet Hamiltonian \(\smash{\tilde{H}(k)}\) to leading order in \(\lambda\) and \(\delta_{b}(k)\) is
\begin{gather}\label{eq:detuned_effective_generator}
\begin{aligned}
    \tilde{H}(k)
    ={}&\frac{\delta_b(k)}{\mathcal{T}_m}\tau_z\\[3pt]
    &+\frac{\lambda}{\mathcal{T}_m}
    \begin{bmatrix}
        0 & \mathcal{S}(\omega(k_b),\mathcal{T}_m)\\
        \mathcal{S}^*(\omega(k_b),\mathcal{T}_m) & 0
    \end{bmatrix}.
\end{aligned}
\end{gather}
After removing the common free-evolution phase at the crossing, \(\smash{\tilde{H}(k)}\) gives the one-word propagator as
\begin{gather}
    \exp(i\epsilon(k_b)\mathcal{T}_m) \mathcal{U}(\mathcal{T}_m,k) \simeq \exp(-i\smash{\tilde{H}(k)}\mathcal{T}_m).
\end{gather}
The eigenvalues of \(\tilde{H}(k)\) are \(\pm R(k)\), where
\begin{gather}
    R(k)=\frac{\sqrt{\delta_b(k)^2+g_b^2}}{\mathcal{T}_m},
    \quad
    g_b=|\lambda|\, |\mathcal{S}(\omega(k_b),\mathcal{T}_m)|.
\end{gather}
Here, \(g_b\) is the dimensionless coupling over one word.
When the first-order expansion is valid and \(|\delta_b(k)|,g_b\ll1\), the quasienergy splitting is \(\varDelta{E}(k)\simeq2R(k)\).
The population at word ends after \(r\) repetition is
\begin{gather}
    P(r\mathcal{T}_m,k)
    \simeq
    \frac{g_b^2}{\delta_b(k)^2+g_b^2}
    \sin^2\!\left[r\mathcal{T}_mR(k)\right].
\end{gather}
Increasing the detuning magnitude raises the oscillation frequency \(2R(k)\) and reduces its amplitude.
For \(g_b>0\), the amplitude falls to half its resonant value at \(|\delta_b(k)|=g_b\).
The corresponding half-width in the free-band separation \(\omega(k)\) is thus \(2g_b/\mathcal{T}_m\simeq\varDelta{E}_b(m)\).
 
% APPENDIX : SECTION G
\section{Detuning from the special momenta}
\label{app:detuning_half_integer}

% 1 Free evolution near k_* :-
Here we derive the population-error bound used in Sec.~\ref{sec:half_integer}.
For \(|\epsilon(k)|T=(s+1/2)\pi+\delta\), the free evolution is \(F(k)=F(k_*)\exp[i\delta\tau_z]\), since \(\epsilon(k)\leq0\).
In the free-band basis, the kick operators are independent of \(k\), so detuning changes only the free steps.
% 2 First-order amplitude :-
The effect on the kick phases is visible at first order in \(\lambda\), where the interband amplitude has magnitude \(\lambda|\mathcal{S}(\omega(k),t_n)|\).
Since \(\omega(k)T = (2s+1)\pi+2\delta\), the Fourier sum \(\mathcal{S}\) becomes
\begin{gather}
    \mathcal{S}(\omega(k),t_n)
    =\sum_{j=1}^{n}(-1)^j f_j\exp(-2i\delta j).
\end{gather}
At \(\delta=0\), this equals \((-1)^n\mathfrak{D}(t_n)\) [Eq.~\eqref{eq:alternating_sum}].
The phase \(2\delta j\) accumulates to order unity over \(n\sim1/|\delta|\) steps.

% 3 Exact bound :-
We now bound the population error without expanding
in kick strength.
The difference between the free evolution operators satisfies
\begin{gather}
    \|F(k)-F(k_*)\| = 2|\sin(\delta/2)| \leq |\delta|,
\end{gather}
where \(\|\cdot\|\) denotes the operator norm, which for a diagonal matrix is the largest absolute value of its diagonal entries.
Multiplication by the same unitary kick preserves this norm, so the one-step operators differ by at most \(|\delta|\).
Since unitary evolution does not amplify the accumulated error, after \(n\) steps we have
\begin{gather}
    \|\mathcal{U}(t_n,k)-\mathcal{U}(t_n,k_*)\| \leq n|\delta|.
    %\|\mathcal{U}(t_n,k)-\mathcal{U}(t_n,k_*)\| \leq\sum_{j=1}^{n}\|U_j(k)-U_j(k_*)\| &\leq n|\delta|.
\end{gather}
In the free-band basis, the transition amplitude is an off-diagonal matrix element of \(\mathcal{U}\), so its difference is bounded by the same operator norm.
For amplitudes of magnitude at most one, their squared magnitudes differ by at most twice their difference.
Thus,
\begin{gather}\label{eq:detuning_bound}
    |P(t_n,k)-P(t_n,k_*)| \leq 2n|\delta|.
    % |P(t_n,k)-P(t_n,k_*)| \leq 2\|\mathcal{U}(t_n,k)-\mathcal{U}(t_n,k_*)\| \leq 2n|\delta|
\end{gather}
This bound holds at any kick strength.

% 4 Window :-
To keep the population error small through \(n=\tau_{\rm F}/T\), it is sufficient to require \(2(\tau_{\rm F}/T)|\delta|\ll1\).
This gives Eq.~\eqref{eq:detuning_requirement}.
The bound guarantees that the population stays close to the resonant trajectory over this interval.
It does not determine a necessary resonance width or the change in the time to first reach \(\mathcal{P}\).

% 5 Class I words :-
For a fixed Class~I word, \(|\mathfrak{D}(\mathcal{T}_m)|=1\).
The weak-kick coupling accumulated over one word is therefore \(\lambda\), while the detuning accumulated over the word
is \(L_m\delta\).
The leading oscillation amplitude from Appendix~\ref{app:detuning} is
\begin{gather*}
    \frac{\lambda^2}{\lambda^2+(L_m\delta)^2}.
\end{gather*}
It falls to half its resonant value when \(|L_m\delta|=\lambda\).
The corresponding half-width is therefore
\begin{gather*}
    |\delta|\simeq\frac{\lambda}{L_m}
    \simeq
    \arcsin\!\sqrt{\mathcal{P}}\,
    \frac{T}{\tau_{\rm I}},
\end{gather*}
where the last relation follows from Eq.~\eqref{eq:periodic_waiting_time}.

\bibliography{ref}

\end{document}